\documentclass[seceq]{ptptex}

\usepackage{graphicx}
\usepackage{wrapft}

\title{
Precision Spectroscopy of Deeply Bound Pionic Atoms and  
Partial Restoration of Chiral Symmetry in Medium}

\author{
Natsumi \textsc{Ikeno}$^{1}$,
Rie \textsc{Kimura}$^{1}$, 
Junko \textsc{Yamagata-Sekihara}$^{2, 3}$, 
Hideko \textsc{Nagahiro}$^{1}$, 
Daisuke \textsc{Jido}$^{3}$, 
Kenta \textsc{Itahashi}$^{4}$,  
Li Sheng \textsc{Geng}$^{5, 6}$
and Satoru \textsc{Hirenzaki}$^{1}$
}

\inst{
$^1$Department of Physics, Nara Woman's University, Nara 630-8506, Japan\\ 
$^2$Departamento de F$\acute{i}$sica Te$\acute{o}$rica and
IFIC, Centro Mixto Universidad de Valencia-CSIC, Institutos de
Investigaci$\acute{o}$n de Paterna, Aptdo. 22085, 46071 Valencia, Spain\\
$^3$Yukawa Institute for Theoretical Physics, Kyoto University, Kyoto 606-8502, Japan\\ 
$^4$Nishina Center for Accelerator-Based Science, RIKEN, 2-1 Hirosawa,
Wako, 351-0198 Saitama, Japan\\
$^5$School of Physics and Nuclear Energy Engineering, Beihang University, Beijing 100191, China\\
$^6$Physik Department, Technische Universit\" at M\"unchen, D-85747 Garching, Germany
}

\abst{
We study theoretically the formation spectra of deeply bound pionic
atoms expected to be observed by experiments with high energy resolution 
at RIBF/RIKEN, and we discuss in detail the possibilities to extract new
information on the pion properties at finite density from the
observed spectra, which may provide
information on partial restoration of chiral symmetry in medium.
We find that the non-yrast pionic states such as 2$s$ are expected to be
seen in
the ($d$,$^3$He) spectra, which will be helpful to reduce uncertainties
of the theoretical calculations in the neutron wave functions in nucleus.
The observation of the 2$s$ state with the ground 1$s$ state is also
helpful to reduce the experimental uncertainties associated in the
calibration of the absolute excitation energy.
We find that the nuclear densities probed by
atomic pions are quite stable and almost constant for various atomic
states and various nuclei.
Effects of the pion wave function renormalization to the formation
spectra are also evaluated.}

\begin{document}

\maketitle

\section{Introduction}
Properties of hadrons at finite density and temperature are extremely
interesting in the contemporary hadron-nuclear physics because they
provide important hints to explore the relation between the symmetry
breaking pattern of QCD and the observed hadron properties for the study
of the QCD vacuum structure.~\cite{RefJ14} \ 
One of the good systems to observe in-medium hadron properties is the
deeply bound pionic atom~\cite{RefJ4,RefJ5}, 
which is a $\pi^-$ atomic bound state hardly
observed by X-ray spectroscopy,
such as the $1s$ or $2p$ states in heavy nuclei. 
The deeply bound states have been experimentally produced in the ($d,^3$He)
reactions with Pb and
Sn isotope targets at GSI~\cite{RefJ1,RefJ2,RefJ3,Geissel} by following
theoretical
predictions.~\cite{RefJ6, RefJ7, RefJ9, RefJ10, RefJ11} \ 
In the latest experiment~\cite{RefJ1}, the energy
shifts and widths of the 1$s$ states have been precisely measured in three Sn
isotopes and isospin-density dependence of the $s$-wave pion-nucleus
potential has been deduced. From these observations, reduction of the
chiral order parameter $\langle \bar q q \rangle$ in nucleus was
concluded. 
A recent model independent theoretical analysis supported the
way to extract the in-medium quark condensate from the pionic atom data
and showed a relation connecting the in-medium quark condensate to the
hadronic observables.~\cite{Jido} \  
For further studies of in-medium pion properties,
new experiments were proposed 
to make high precision spectroscopy of pionic atoms systematically 
in RIBF/RIKEN.~\cite{RIBF1,RIBF2} 
Thus, the pionic atom can be one of the best systems to deduce the
quantitative results for the meson properties and the partial restoration
of chiral symmetry in medium around normal nuclear density at zero temperature.

Based on these theoretical and experimental developments,
we think that we should consider now the possible future directions of 
the studies of the deeply bound pionic atoms  after
15 years from the discovery of the deeply bound pionic atoms at
GSI.~\cite{RefJ2,RefJ3,Yamazaki,Z.Phys.A356} \ 
We discuss the following points in this article in detail,
\begin{itemize}
\item[(i)]  Predictions of the pionic atom formation spectra by the ($d, 
^{3}$He) reactions on Sn and Te isotope targets 
proposed in Refs.~\citen{RIBF1,RIBF2} at RIBF/RIKEN,
\item[(ii)] 
Possibility to observe pion properties and to determine the value of the
chiral condensate at different nuclear density from 
$\rho=0.6\rho_{0}$ to obtain information on 
the symmetry breaking parameters beyond the linear density 
approximation,
\item[(iii)]
Possibility to determine the wave function renormalization factor 
introduced in Refs.~\citen{Jido} and \citen{Kolo}  
from pionic atom observables,
\item[(iv)]
Uncertainties included in theoretical calculations used to evaluate 
formation cross sections,

\end{itemize}
with paying the attention to the  advantages of the simultaneous
observation of the non-yrast 2$s$ bound state together with the deepest
1$s$ state for the same nucleus in the new experiments with better
energy resolution.~\cite{RIBF1,RIBF2} 

In Section~2, we summarize the theoretical formalism used to connect the pion
properties to the order parameter of the chiral symmetry and the wave
function renormalization factor. 
We also mention the sensitivities of atomic pion to nuclear densities
and the strong correlation between potential parameters.
In Section~3, we show the numerical results and discussions for
the structure and formation of deeply bound pionic atoms. 
Section~4 is devoted to the conclusion.
We summarize the supplementary formalisms and numerical results in Appendixes.

\section{Formalism}
\label{Formalism}
We introduce briefly the formula reported in Ref.~\citen{Jido} in this
section  as the guide
to deduce the information on chiral symmetry from the pionic atoms observables.
We also mention the sensitivities of atomic pions to nuclear density and the
strong correlation of the potential parameters.
The theoretical formula used in this article to calculate the structure and the
formation cross sections of the pionic atoms are summarized
in Appendix~\ref{struct-cross}.

\subsection{Chiral Dynamics of Pionic Atoms} \label{Chiral}
The spontaneously broken chiral symmetry in vacuum is expected to be
restored partially in nuclear medium. 
The partial restoration of chiral
symmetry takes place with effective reduction of the chiral quark
condensate $\langle \bar q q \rangle$ in medium.~\cite{RefJ19,RefJ20} \
Experimental
observation of the reduction of the quark condensate is not so trivial,
since the quark condensate is not a direct observable.
In Ref.~\citen{RefJ1}
K.~Suzuki {\it et al.} made use of in-medium extensions of the two
well-known relations, Gell-Mann-Oakes-Renner (GOR) relation~\cite{Gell} and
Tomozawa-Weinberg relation (TW)~\cite{TW1,TW2}, to connect the quark
condensate and the
observables in pionic atoms, after having extracted the values of the
parameters in the optical potential from the observed energy
shifts and widths of the deeply bound atomic states. 
The $b_1$ parameter of the optical potential in Eq. (\ref{Vopt_Swave1}) 
corresponds to the effective scattering length between pion
and nucleus, which can be expressed by an in-medium extended TW relation:
\begin{equation}
T^{(-)}_{\pi A} = - 4\pi \epsilon_1 b_1 = \frac{\omega}{2 f_\pi^{*2}},
\label{T_piA}
\end{equation}
where $f_\pi^*$ is an in-medium pion decay constant. 
Assuming that the pion mass does not change in nucleus, one could have
an in-medium GOR relation 
\begin{equation}
m_\pi^2 f^{*2}_\pi = - 2 m_q \langle \bar qq \rangle_\rho,
\end{equation}
where $m_q$ is the isospin-averaged quark mass $m_q=(m_u+m_d)/2$ and
$\langle \bar qq \rangle_\rho$ is the in-medium quark condensate. Using
these two relations, K.~Suzuki {\it et al.}~\cite{RefJ1} could get a connection
between the experimental observation and the in-medium quark condensate
through $b_1$ and $f_\pi^*$. 

The experimental observations of the deeply bound pionic states and the
attempt to connect the observables to $\langle \bar qq \rangle_\rho$
based on the simple extension of the in-vacuum relations have stimulated
theoretical works to give stronger foundations of the analysis. It was
shown based on chiral perturbation theory~\cite{Kolo} and on a correlation
function analysis~\cite{Jido} that the in-medium TW relation
Eq.~(\ref{T_piA}) can be valid
in the linear approximation of the isovector density, which leads to
\begin{equation}
\frac{b_1^{\rm free}}{b_1} = \left( \frac{f_\pi^t}{f_\pi} \right)^2,
\label{b1-fpi}
\end{equation}
with the in-vacuum isovector $\pi N$ scattering length $b_1^{\rm free}$,
the in-vacuum pion decay constant $f_\pi$ and the time component of the
in-medium pion decay constant $f^t_\pi$. 
Further, it was found in a
model-independent argument based on the operator relation that there is
a sum rule for the in-medium quark condensate.~\cite{Jido} \ 
This sum rule can be
simplified at low density limit and gives a new scaling relation
\begin{equation}
 \frac{\langle {\bar q}q\rangle_{\rho}}{\langle {\bar q}q\rangle}
=Z^{*1/2}_{\pi} \left(\frac{f^{t}_{\pi}}{f_{\pi}}\right),
\label{qqbar}
\end{equation}
where $Z_\pi^*$ is the wave function renormalization for the in-medium
pion (see also Ref.~\citen{Jido2}). 
The density dependence of the wave function renormalization can be
estimated at low density limit through the $\pi N$ scattering
amplitude~\cite{Jido} :
\begin{equation}
Z^{*}_\pi = 1 - \beta \rho,
\label{Zpi}
\end{equation}
with $\beta = 2.17 \pm 0.04$ fm$^3$. 
Combining Eqs.~(\ref{b1-fpi}) and (\ref{qqbar}), one obtains
a connection between the in-medium quark condensate and the experimental
observables as
\begin{equation}
 \frac{\langle {\bar q}q\rangle_{\rho}}{\langle {\bar q}q\rangle}
=Z^{*1/2}_{\pi} \left(\frac{b^{\rm free}_{1}}{b_{1}}\right)^{1/2}.
\label{qqbar2}
\end{equation}
The $b_1$ parameter in Eq.~(\ref{qqbar2}) is obtained by the pionic atom
data, and the rest in the right hand side, $b_1^{\rm free}$ and
$Z^*_{\pi}$, are evaluated by the $\pi N$ scattering.
For completeness, it is very good if one can determine also the wave
function renormalization from the deeply bound pionic atoms instead of
using the in-vacuum $\pi N$ scattering. 
We discuss the possibility to determine the pion wave function
renormalization factor from the observation of the pionic atom states in
Section~3.3. 
We also mention here that the estimation of the higher order effects of
the density from experimental data is also interesting and important to
explore the behavior of $\langle {\bar q}q\rangle_{\rho}$ beyond the linear
density approximation.~\cite{Doring,Jido21,Jido22} \ 
The discussion of the
possibility to deduce the higher order effects from the observation will
be given in Section~\ref{Sec.3.2}.

\subsection{Nuclear densities probed by atomic pion and Seki-Masutani correlations} \label{Seki-Masutani}
As we will see in detail in Section~\ref{Sec.3.2}, the pion in
atomic states observed by the X-ray spectroscopy are known
to be only sensitive to narrow range of nuclear density, which is almost
independent of the nuclides and the atomic states.~\cite{rho_e} \ 
Hence, the structure of the pionic atoms is essentially determined by 
the optical potential strength at the effective nuclear density
$\rho_e$ probed by atomic pion.
Namely, the series of optical potentials which have the same potential
strength at $\rho=\rho_e$ equivalently provide the almost same structure
of the pionic atoms.
As a consequence, we have a certain relation between potential
parameters of the optical potentials which reproduce the atomic data well.
We consider the $s$-wave part of the optical potential $V_{s}$ as
an example which includes the important piece $b_{1}$ to
deduce the $\langle {\bar q} q \rangle_\rho$ value and plays the dominant
role for the deeply bound $s$ states.~\cite{RefJ10} \ 
The $s$-wave optical potential for the symmetric nuclei $\rho_n =
\rho_p$ is written as,
\begin{eqnarray}
2\mu V_{s}(r)
&=&-4\pi[ \varepsilon_{1} b_{0}\rho(r)+\varepsilon_2B_0\rho^2(r)].
\label{SM4}
\end{eqnarray}
For all $b_0$ and $B_0$ values satisfying the relation
\begin{equation}
b_0 + \frac{\varepsilon_2}{\varepsilon_1} \rho_e B_0 = {\rm constant}, 
\label{SM*}
\end{equation}
the optical potential~(\ref{SM4}) has the same strength at $\rho=\rho_e$ and provide 
almost same structure of the pionic atoms. 
Actually the correlations between potential parameters are found
phenomenologically and called as the Seki-Masutani (SM) correlations.~\cite{SM} \ 

The SM correlations are expressed as;
\begin{eqnarray}
b_0+ \alpha_{s} B_0 = \beta_s = (-0.03+0.01 i)\rm{m}_\pi^{-1}, 
\hspace{5mm}
\alpha_{s} \simeq 0.23 m_{\pi}^3,
\label{SM1}
\end{eqnarray}
and
\begin{equation}
\frac{c_0+ \alpha_{p} C_0}{\gamma} = \beta_p
=(-0.2+0.02 i)\rm{m}_\pi^{-3}, 
\label{SM2}
\hspace{5mm}
\alpha_{p} \simeq 0.37 m_{\pi}^3,
\end{equation}
where,
\begin{equation}
\gamma = 1+\frac{4}{3} \pi \lambda c_0 \rho_e.
\label{SM3}
\end{equation}
The parameters $b_{0}, B_{0}, c_{0}$, and $C_{0}$ determine the strength
of the pion-nucleus optical potential shown in  Eq.~(\ref{Vopt}). 
Seki and Masutani found that the series of potential parameter sets, which
satisfy the relations (\ref{SM1}) and (\ref{SM2}), reproduce the observed
data by the X-ray spectroscopy reasonably well.
Namely, it is very difficult to fix the unique set of 
potential parameters from the data taken by the X-ray experiments.
In the modern $\chi^{2}$ analyses of the data, one can expect the unique
determination of the potential parameters and, actually, one may find
literatures which report the unique determination.
However, there still remain the strong SM correlations, and the
$\chi^{2}$-values in these analyses have the long deep valley structure
along the SM relations of potential parameters.

We can find from Eqs.~(\ref{SM*}) and (\ref{SM1}) as,
\begin{eqnarray}
\rho_e = \frac{\varepsilon_1}{\varepsilon_2} \alpha_s,   
\label{SM6}
\end{eqnarray}
and the value of $\alpha_{s}$ determined by the analyses of atomic data
indicates, 
\begin{equation}
\rho_e \simeq \displaystyle \frac{1}{2}\rho_0.
\label{SM7}
\end{equation}
Hence, it means that the SM correlation found phenomenologically is the
consequence of the fact that the
nuclear density probed by various pionic atoms observed by X-ray is
always close to $\displaystyle \frac{1}{2}\rho_{0}$.
This feature is also found for some of the deeply bound pionic states
\cite{RefJ5,rho_e}, and thought to be one of the robust features of
pionic states.
In the early stage of the exploration of deeply bound pionic atoms, 
this robustness was an advantage
to deduce the pion properties at $\rho_e \simeq \displaystyle \frac{1}{2}
\rho_{0}$,
however, to make a step further, we need to investigate the possibility
to deduce the pion properties precisely not only at $\rho_e \simeq \displaystyle
\frac{1}{2} \rho_{0}$ but also at various nuclear densities.
If we can determine $\langle {\bar q} q \rangle_\rho$ for various densities,
it is expected to provide important experimental information for the
studies of the chiral condensate beyond the linear density approximation.

\section{Numerical Results and Discussions}\label{Result}
In this section, we investigate how we can deduce the 
$\langle {\bar q}q\rangle$ value at finite density precisely from
pionic atom data.
As already mentioned, $\langle {\bar q}q\rangle$  was determined in 
Ref.~\citen{RefJ1},
however, the recent theoretical work provide a different formula including
the $Z^{*}_{\pi}$ factor as introduced in Section~\ref{Chiral}.
In addition, the $\langle {\bar q}q\rangle$ value was determined only at 
$\rho \sim 0.6\rho_{0}$
because of the limited sensitivity of the observed pionic atom data to
nuclear density, which is related to Seki-Masutani correlations to the
potential parameters as mention in Section~\ref{Seki-Masutani}.
Thus, we study theoretically these points to find the proper way to
deduce the $\langle {\bar q}q\rangle$ value for various densities from
the future observations.
We also pay attention to the possible simultaneous observation of the
non-yrast 2$s$ state with the deepest 1$s$ state.~\cite{RIBF1,RIBF2}

\subsection{($d, ^{3}$He) spectra in experiments}\label{GSI-Exp}
\begin{figure}[!ht]
\centerline{\includegraphics[width=10.0cm,height=16.5cm]{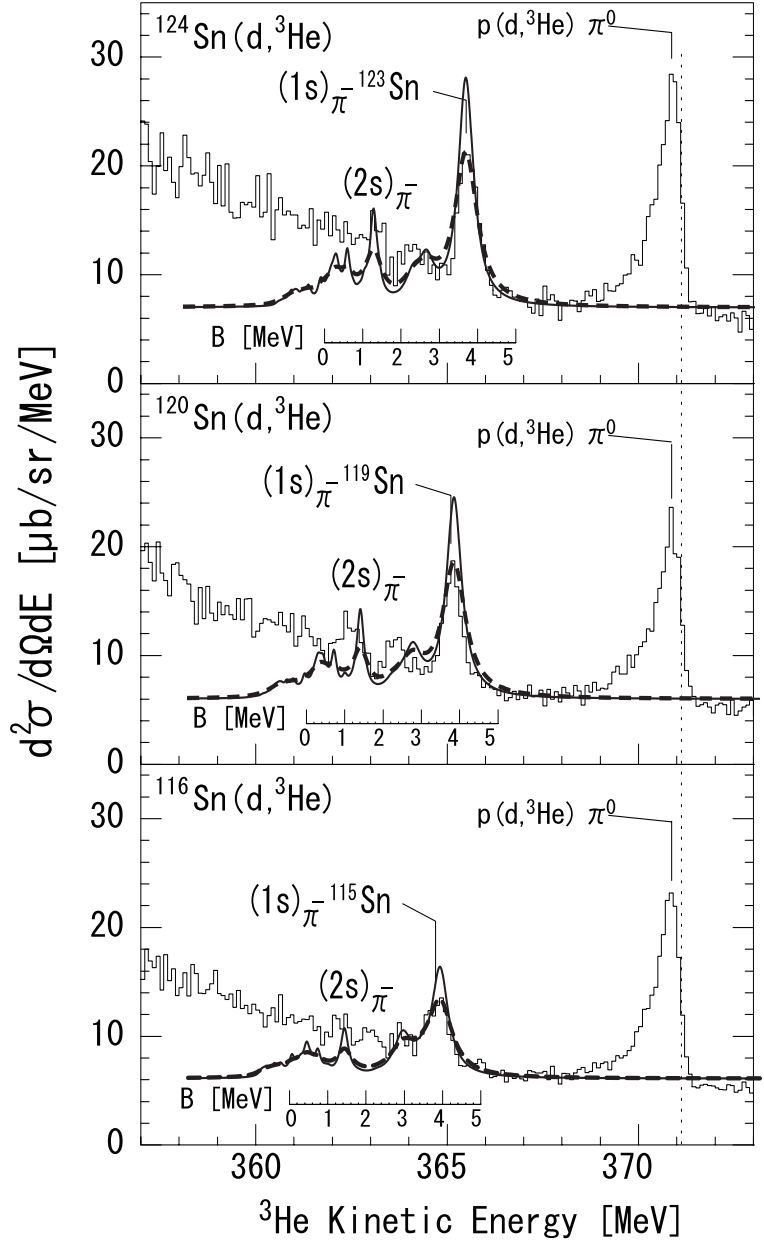}}
\caption{The ($d, ^{3}$He) spectra for the formation of the deeply bond 
pionic states in Sn isotopes as indicated in figure.
The solid and dashed lines denote the theoretical calculations, which
include only bound state
contributions, while the histogram shows observed experimental data~\cite{RefJ1}.
The instrumental energy resolution is assumed to be $\Delta E=394$ keV 
FWHM 
for the dashed lines and 150 keV FWHM for the solid lines.
The calculated results are scaled to reproduce the strength of the 1$s$
state formation peak in the observed spectra for $\Delta E=394$ keV FWHM.}
\label{GSI}
\end{figure}

We first show the comparison of the
theoretical calculations with the latest data of the pionic atom
formation in Sn isotopes.
In Fig.~\ref{GSI}, we show the calculated results together with
experimental data reported in Ref.~\citen{RefJ1}.
The instrumental resolution is assumed to be $\Delta E=394$ keV
FWHM (Full Width Half Maximum) as same as
the data.~\cite{RefJ1} \ 
We added the constant background to the calculated results and then
scaled the calculated spectra to reproduce the strength of the resonance
peak of the 1$s$ state formation in the experimental data for 
this resolution.
We also show the calculated results using the same
theoretical model with the improved resolution
$\Delta E=150$ keV FWHM which is expected to be archived in the planned
experiments.~\cite{RIBF1,RIBF2}. \ 
We find that the theoretical calculation reproduce the experimental data
reasonably well and provide the reliable interpretation of the
spectra, which is an essential foundation of the research activity to deduce
the pion properties from the ($d, ^{3}$He) spectra.
And as we can see from the figure, 
we conclude that the pionic 2$s$ state can be observed as a peak
structure together with the 1$s$ state in the experiments with the better
resolution $\Delta E=150$ keV FWHM.
This result helps to propose the new experimental
activity and motivate us to proceed the 
theoretical analyses reported in this article.

Simultaneous observation of the non-yrast $2s$ state with
the ground $1s$ state is of essential importance in the
experimental viewpoints. We expect smaller experimental 
uncertainty in the energy difference between the two states 
while larger ambiguities are usually associated in the 
calibration of the absolute excitation energy. 

In order to resolve the two states $1s$ and $2s$ in 
the excitation spectrum, the spectral energy 
resolution $\Delta E$ must be improved by a factor of about 
two from $\sim$400 keV to $\sim$200 keV.
The resolution is governed by two major
contributions, namely the incident beam energy 
spread and the energy loss in the target.
The latter can be controlled by changing the target thickness.
However, adaptation of thinner target results
in smaller statistical merits or in longer
data accumulation periods unless the beam
intensity is increased as compensation.

A new experimental approach is thus started~\cite{RIBF2}
in the RI beam factory, 
RIKEN~\cite{RIBF} to achieve improved resolution
of 200 keV with much higher statistics.
The experiment makes full use of the very high 
intensity deuteron beam of $\sim 1 \times 10^{12}$/second, 
which is more than 30 times higher than the previously 
available intensity in the SIS-18 accelerator in the GSI.
Thus, the statistical merit is still huge even if 
the adopted target thickness of 5 mg/cm$^2$ is
factor of three smaller than the previous experiments, 
and the larger luminosity is essential for systematic study over wide
range of nuclei.
The other contribution in the energy resolution,
the incident beam energy spread, is $\sim$ 5 times
larger in the RIBF. Elaborate study on the accelerator,
the beam transfer line, and the spectrometer is ongoing to establish
a dispersion matching beam optics where the contribution
of the beam energy spread is suppressed in the excitation 
spectra~\cite{APR}.

\begin{figure}[t]
\begin{center}
\includegraphics[width=9cm, height=6.5cm]{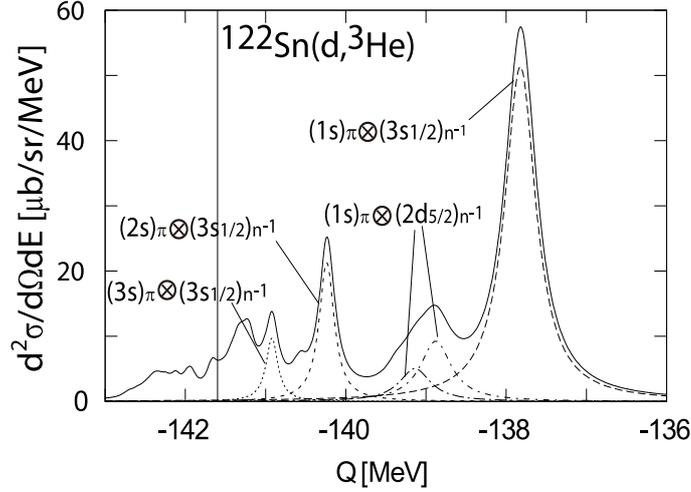}
\caption{Calculated $^{122}$Sn($d, ^{3}$He) spectra for the formation of the
 pionic bound states at $T_{d}=500$MeV are shown as functions of the reaction
 $Q$-value, where the $F_{O}$ factors in Table~\ref{Sn-Fo} are used.
The dominant subcomponents are also shown as indicated in the Figure.
Instrumental energy resolution is assumed to be 150 keV FWHM.
The vertical line indicates the threshold $Q=-141.6$ MeV.}
\label{122Sncross}
\end{center}
\end{figure}

Simultaneous observation of the 2$s$ state with the 1$s$ state is
important also in the theoretical viewpoints.
As shown in detail in Appendix~\ref{AppendixC}, the
uncertainties of the theoretical calculation of the pionic atom
formation cross sections due to the neutron wave function and the
nuclear excited levels could be an obstacle to deduce
the properties of pion by the precision measurements.
Thus, as a practical way to reduce the uncertainties and deduce the
reliable results, we should make use of the recoilless
kinematics to populate the plural number of pionic states coupled with
the same neutron hole state as dominant contributions in the ($d, ^{3}$He) spectra. 
And by comparing the strength of these contributions, we can effectively
remove the ambiguities due to the structure of the target nucleus, the
nuclear excited levels, and the neutron wave functions.
In this context, the simultaneous observation of 1$s$ and 2$s$ pionic
states coupled with the $(3s_{1/2})^{-1}_{n}$ neutron hole is very interesting.
We show the calculated $^{122}$Sn($d, ^{3}$He) spectra in Fig.~\ref{122Sncross}.
This reaction is proposed in Refs.~\citen{RIBF1,RIBF2}.
As we can see form the figure, the 1$s$ and 2$s$ pionic states coupled
with the same $(3s_{1/2})^{-1}_{n}$ neutron hole state can be seen as the
clear peak structures with the realistic energy resolution $\Delta E=150$
keV FWHM.
Thus, the $^{122}$Sn($d, ^{3}$He) reaction can be one of the good reactions
for our purpose.
The expected spectra for Te isotopes are summarized in Appendix~\ref{AppendixD}.

\subsection{Determination of $b_{1}$ parameter at various nuclear density} \label{Sec.3.2}
As shown in Ref.~\citen{RefJ1}, the observed binding energies and widths of
bound pions have been used to determine the strength of the $s$-wave
isovector potential parameter ($b_{1}$).
The important points we should address here are to find out the way to determine $b_1$
value at various densities to know its density dependence beyond the
linear form.

\begin{figure}[t]
\centerline{\includegraphics [width=8.5cm, height=11.0cm]{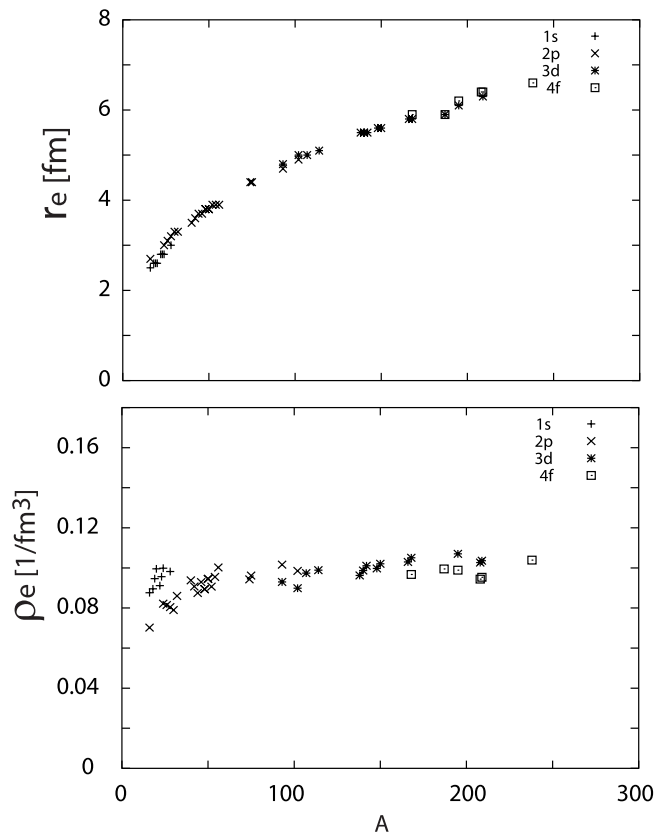}}
\caption{Calculated peak positions $r_{e}$ (upper frame) and the
 corresponding effective nuclear densities $\rho_{e}$ (lower frame) for the  
observed pionic atom states by X-ray experiments are plotted as
functions of nuclear mass number $A$.
The quantum numbers of each atomic state are indicated in the figure.}
\label{fig:5}
\end{figure}

\begin{figure}[t]
\begin{center}
\includegraphics [width=6.5cm, height=8.5cm]{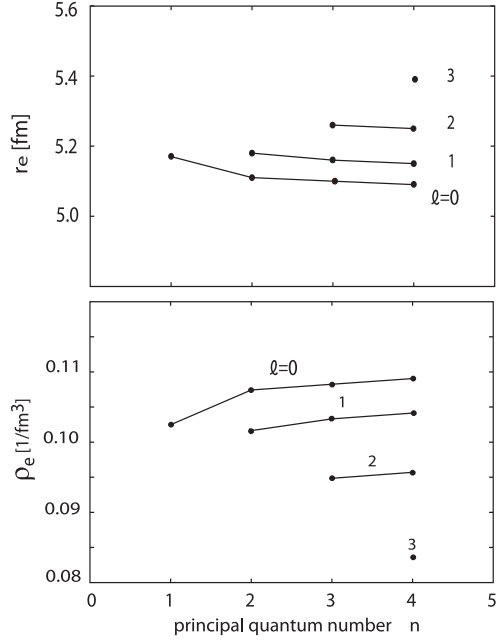}
\caption{Peak positions $r_{e}$ (upper frame) of the overlapping
densities $S(r)$ and the corresponding effective nuclear densities
$\rho_{e}$ (lower frame) of $^{121}$Sn.}
\label{fig:4}
\end{center}
\end{figure}

\begin{figure}[t]
\centerline{\includegraphics [width=10.0cm, height=10.5cm]{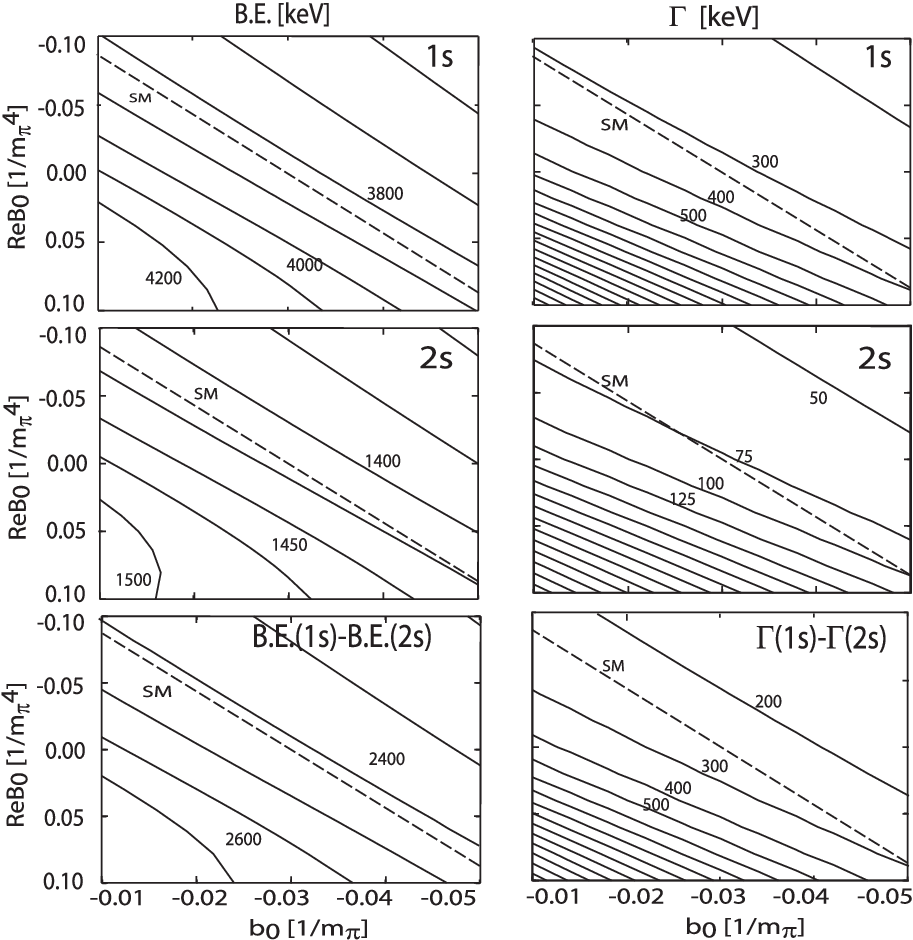}}
\caption{Contour plots of the binding energies (left) and widths
(right) for the 1$s$ (upper) and 2$s$ (middle) states, and  
the difference of 1$s$ and 2$s$ states (lower) in $^{121}$Sn in the
$b_0$-Re($B_0$) plane.  
The dashed lines depicted by SM are the parameter sets satisfying the
Seki and Masutani correlation.~\cite{SM} \ 
The numbers in the figure indicate the values of the binding energy and
 width of contour lines in unit of keV.}
\label{contour-b0}
\end{figure}

\begin{figure}[htb]
\begin{center}
\centerline{\includegraphics [width=10.0cm, height=10.5cm]{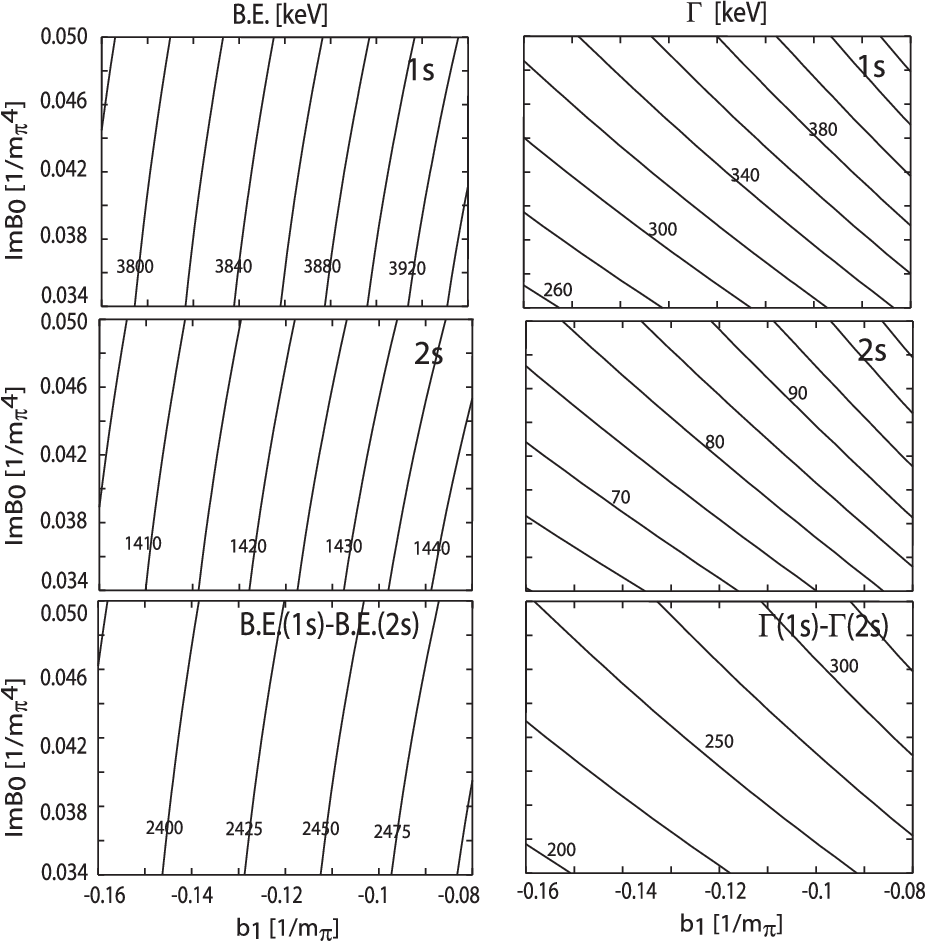}}
\caption{Contour plots of the binding energies (left) and widths
(right) for the 1$s$ (upper) and 2$s$ (middle) states, and 
the difference of 1$s$ and 2$s$ states (lower) in $^{121}$Sn 
in the $b_1$-Im($B_0$) plane.
The numbers in the figure indicate the values of the binding energy and
 width of contour lines in unit of keV.}
\label{contour-b1}
\end{center}
\end{figure}

First, we consider the effective nuclear density probed by atomic
pion.~\cite{rho_e} \ 
The effective nuclear density $\rho_{e}$ is defined as the nuclear density
at the radial coordinate $r_{e}$ as $\rho_{e}=\rho(r_{e})$, 
where the overlapping density,  
\begin{equation}
S(r)=\rho(r)\mid R_{nl}(r) \mid^2r^2, 
\label{overlap}
\end{equation}
has the maximum value.
Here, $R_{nl}(r)$ is the radial wave function of the pionic atom in a 
state of ($nl$).
The definition of the overlapping density $S(r)$ indicates that it is
expected to evaluate the sensitivity of the iso-scalar $s$-wave optical
potential term to the energy eigen values in the sense of the first order
perturbation theory.
In Fig.~\ref{fig:5}, we show the calculated $r_{e}$ and $\rho_{e}$
systematically for the pionic atom states which have been observed by
the experiments of the X-ray spectroscopy, so far.\cite{Xray} \ 
As shown in the figure, $r_{e}$ increased with the nuclear mass
number $A$ monotonically, however $\rho_{e}$ is almost constant
(`saturated') except for the very light nuclear cases.
These features, which are almost independent on the quantum numbers of
the states, can be understood by the pocket structure of the potential for
the atomic pion and the localization of the overlapping densities there
as can be seen in Appendix~\ref{AppendixB}.
Thus, all pionic atoms observed by the X-ray experiments probed the
almost same nuclear density nearly independent on the nuclide and the
quantum numbers of the pion except for the pionic states in the light nuclei.
Thus, it seems that all observations of these states only provide pion
properties around 0.6$\rho_0$.
To obtain pion properties at various nuclear densities, it will be
necessary to observe other states.

Then, we consider the atomic states in $^{121}$Sn up to $n=6$ including
the deeply bound and non-yrast states, which can not be observed in
X-ray experiments.
When we use the Coulomb wave functions for bound pions,
$\rho_{e}$ distributed in relatively wide range $0.05
\lesssim \rho_{e} \lesssim 0.16$ [fm$^{-3}$] depending on the bound states.
As naturally expected, the states with smaller orbital angular momentum
$\ell$ tend to probe the larger nuclear densities.
As the realistic cases, we show the results calculated with the optical
potential in Fig.~\ref{fig:4}. 
$\rho_{e}$ and $r_{e}$ only
change inside the smaller range than those calculated with Coulomb
potential only. 
This behavior can be also understood by the picture of the potential
pocket and the localization of the overlapping densities at the nuclear
surface.
Thus, we find that the effective nuclear density is rather stable
($0.08 \lesssim \rho_{e} \lesssim 0.11$ [fm$^{-3}$])
even for the non-yrast states and the deeply bound states.

We mention here that an interesting tendency appeared in
Fig.~\ref{fig:4} that the states with higher
$n$ for a fixed $\ell$ provide larger $\rho_{e}$ values contrast to the
usual intuitions.
This is because of the larger tunneling effects to the central 
soft core for more lightly bound states.
Thus, the sensitivity of pion in the higher $n$ states moves to higher
$\rho$, however at the same time, the absolute magnitude of the strong
interaction effects are reduced rapidly as $n$ increases.
For example, 
we may think that it is better to observe the 4$s$ and 4$f$ states to
probe pion properties in different nuclear densities from Fig.~\ref{fig:4}.
This will be wrong since the strong interaction effects are
too small for both states and the $p$-wave interaction plays dominant
role in $f$ states~\cite{RefJ10} and hides the $b_{1}$ effects even they may provide
information for different $\rho$.

Hence, we find that the observation of the binding energies of the
pionic atoms will provide the pion properties near $\rho \sim 0.6
\rho_{0}$.
Thus, to deduce the density dependence of the $b_{1}$ parameter, we
require the extremely higher energy resolution data as we will see later
in the discussion with energy contour plots in this section.

We, then, consider the contour plots of the binding energies B.E. and
widths $\Gamma$ of 1$s$ and 2$s$ pionic states in $\pi^{-}-$$^{121}$Sn
system.
By the contour plots of eigen energies, we can see the difference of the
sensitivities to the nuclear density as the deviations from the SM
correlation and we can see the required energy resolution to distinguish
the pion
properties at different $\rho$.
The contour plots of the differences of the binding
energy and width of 1$s$ and 2$s$ states are also shown, which could be used
to deduce the systematic errors due to the calibration of the absolute
binding energies and the uncertainties of the neutron distribution of
$^{121}$Sn.
We mention here that the binding energies and widths of the 1$s$ states
in Sn isotopes were precisely determined in the last
experiment.~\cite{RefJ1} \ 
For example, the binding energy and width of the 1$s$ state in
$^{123}$Sn are B.E.$=3.744 \pm 0.018$ [MeV] and $\Gamma= 0.341 \pm 0.072$
[MeV]~\cite{RefJ1}.

We show the numerical results in Fig.~\ref{contour-b0} in $b_{0}-$Re$B_{0}$
plane together with the SM correlation line Eq.~(\ref{SM1}).
We find that the contour lines for the binding energy of 1$s$ state are
almost parallel to the SM line.
Since the slope of the contour lines in $b_{0}$-Re$B_{0}$ plane provide the
information of the nuclear density probed by the atomic pion as shown in
Eqs.~(\ref{SM1}) and (\ref{SM6}), this feature 
indicates that the nuclear density mainly probed by the 1$s$ state is
very close to $\alpha_{s}$ as shown in
Eq.~(\ref{SM6}), and that the precise measurements of the
1$s$ state provide the medium effects of pion at $\alpha_{s}$.
In addition, it is very hard to determine the unique parameter set
($b_{0}$, Re$B_{0}$) on the SM line only from the observation of 1$s$ state.

On the other hand, the contour lines of the 2$s$ state show the slightly
different slope from that of the SM line.
This behavior indicates that the properties of the 2$s$ state are determined by the
pion properties at slightly different nuclear density from $\alpha_{s}$
as indicated by the result shown in Fig~\ref{fig:4}.
And we may be able to find the unique parameter set ($b_{0}$, Re$B_{0}$)
by using the both B.E. of 1$s$ and 2$s$ states with high precision.
The plots of the difference of the binding energies of 1$s$ and 
2$s$ states, which is expected be observed with high precision without
systematic errors due to the absolute energy calibration,  
show the similar behavior of that of the 1$s$ state because
the larger binding energy value of the 1$s$ states dominate the behavior of
the plot.
We may also expect that the uncertainties due to the neutron
distribution are partly cancelled in the plots of the energy
differences.

In principle, to determine the $b_{1}$ value for different $\rho$ by
observing the atomic states, we need to distinguish
the nuclear density observed by these states, which will be
equivalent to fixing the unique set of \{$b_{0}$, Re$B_{0}$\} 
by data, and then to determine $b_{1}$ value
independently for each state.
Thus, the experimental data with very high precision are
necessary to obtain $b_1$ values at various $\rho$ as indicated in
Fig.~\ref{contour-b0}.

The contour plots of the widths show different behavior from those
of the binding energies, and the contour lines are not parallel to 
the SM line for all three cases shown in the right panels in
Fig.~\ref{contour-b0}. 
To understand this behavior intuitively is a little difficult 
since this is the effect to the imaginary eigenvalues from the
modifications of the real part of the potential.
However, we can naively expect that the $\rho^{2}$ behavior of the
imaginary potential can provide the different sensitivity of the widths
of the pionic states from that of the binding energies.
The contour plots of widths indicate that the precise determinations of the
pionic widths will provide the constraints to the potential
parameters, though the determinations of the widths are more difficult in
general than those of the binding energies.

In Fig.~\ref{contour-b1}, we show another contour plot in
$b_{1}-$Im$B_{0}$ plane.
The $b_{1}$ parameter is one of the most important parameters and
has the very close relation to $\langle {\bar q} q \rangle$, and the
Im$B_{0}$ parameter is the leading term to determine the width $\Gamma$
in the $s$-states.
The contour of the binding energy indicates the reasonable independence of
B.E. on Im$B_{0}$, while the contour of the width indicates the
complexity of the behavior of the width which depends strongly both on
Im$B_{0}$ and $b_{1}$.
The contour of the binding energy shows a clear relation between the
accuracy of the binding energy data and the $b_1$ parameter determination.

One possible way to deduce the $b_{1}$ values for different $\rho$ is the
parameter search using the most precise experimental data of the atomic $s$
states, which are most sensitive to the $s$ wave potential parameters~\cite{RefJ10}, 
allowing the different value of $b_{1}$ for each states.
Since this procedure, in principle, require to distinguish the nuclear densities
observed by the atomic states to determine the different $b_{1}$
values for these states independently, 
the precisions of the data should be so high that we can
clearly determine the $b_0$ and $B_0$ parameter uniquely and, thus, 
the feasibility of this procedure highly depends on the
precision of the pionic atom data.

\subsection{($d, ^3$He) spectra and observation of $Z^{*}_{\pi}$ }\label{Sec.3.3}
The purpose of this section is to investigate the possibilities to
determine the wave function renormalization factor $Z^{*}_{\pi}$ shown in
Section~\ref{Chiral} from
the observation of the formation cross section of the deeply bound
pionic atoms.
The basic idea is to observe the change of the cross section due to the
modification of the pion wave function $\phi_{l_{\pi}}$ as, 
\begin{equation} 
\phi_{l_{\pi}}(\boldsymbol{r}) \rightarrow \phi^{R}_{l_{\pi}}(\boldsymbol{r})=Z^{*1/2}_{\pi}
\phi_{l_{\pi}}(\boldsymbol{r}),
\label{WF_ren}
\end{equation}
due to the renormalization factor $Z^{*}_{\pi}$ originated from the
energy dependence of the pion selfenergy.
The modification of the wave function causes the change of the 
the cross section as shown in Appendix~\ref{struct-cross}.
$Z^{*}_{\pi}$ is associated with the
pion selfenergy in nuclear medium and, hence, has the $\rho(r)$
dependence as shown in Eq.~(\ref{Zpi}) in contrast to that appeared in the
standard text book of filed theory.
Due to the $\rho(r)$ dependence of $Z^{*}_{\pi}$, we can expect to have
different effects of $Z^{*}_{\pi}$ for the formation cross sections of
different subcomponents [$\ell_\pi \otimes j^{-1}_n$] in general.

\begin{figure}[t]
\begin{center}
\includegraphics [width=7cm, height=8.5cm]{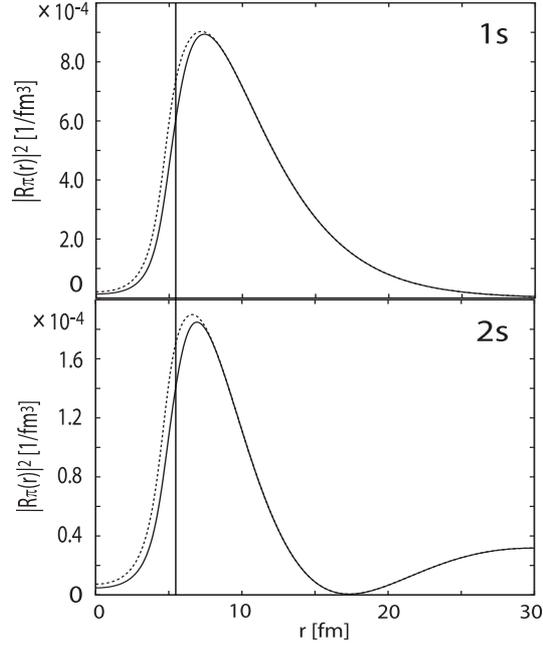}
\caption{Pion radial density distributions $|R_{\pi}(r)|^2$ are shown as
functions of the radial coordinate $r$ for 1$s$ and 2$s$ states in
$^{121}$Sn. Solid and dotted lines show the pion density evaluated by
the radial part of the renormalized ($\phi^{R}$) and unrenormalized
($\phi$) wave functions defined in Eq.~(\ref{WF_ren}), respectively.
The vertical line indicates the nuclear half radius $R=5.4761$ fm.}    
\label{WF(ren)}
\end{center}
\end{figure}

\begin{figure}[t]
\begin{center}
\includegraphics [width=7.3cm, height=5.7cm]{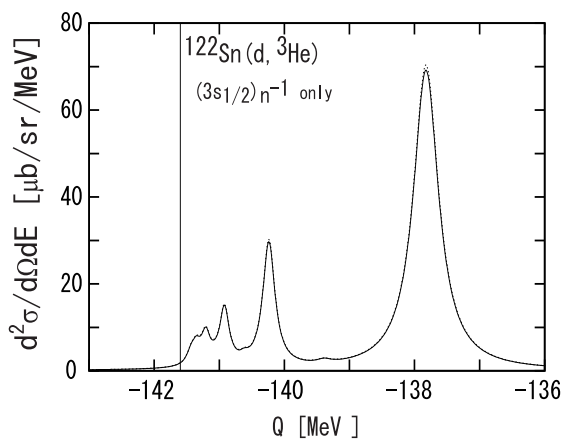}
\caption{Contributions of the $(3s_{1/2})^{-1}_{n}$ neutron hole state 
to the $^{122}$Sn($d, ^{3}$He) spectra for the
formation of the pionic bound states are shown as functions of
the reaction $Q$-value.
The $F_{O}$ and $F_{R}$ factors are fixed to be 1.
The solid line shows the calculated results with the renormalized pion
wave function ($\phi^{R}$) and the dotted line that with the
unrenormalized wave function ($\phi$) defined in Eq.~(\ref{WF_ren}).
The vertical line indicates the threshold $Q=-141.6$ MeV.}
\label{Cross-ren}
\end{center}
\end{figure}

\begin{table}[t]
\begin{center}
\caption{The calculated effective numbers of $[(1s)_{\pi}\otimes
 (3s_{1/2})^{-1}_{n}]$ and $[(2s)_{\pi} \otimes (3s_{1/2})^{-1}_{n}]$ 
subcomponents by renormalized and unrenormalized
 pion wave functions defined in Eq.~(\ref{WF_ren}) in PWIA.
The ratios of the effective numbers of 1$s$ and 2$s$ pionic states are also shown.}
\begin{tabular}{cccc} \\ \hline\hline 
pion wave function & $N_{\rm eff}(1s)$ & $N_{\rm eff}(2s)$ &$N_{\rm
 eff}(1s)/N_{\rm eff}(2s)$  \\ \hline
$\phi^{R}_{l_\pi}(r)$ & $1.36 \times 10^{-1}$  & $2.92 \times 10^{-2}$ & 4.66 \\ 
$\phi_{l_\pi}(r)$ & $1.41 \times 10^{-1} $ & $3.00 \times 10^{-2}$ &  4.70  \\\hline
\label{Neff(D=1)}
\end{tabular}
\end{center}

\begin{center}
\caption{Same as Table~\ref{Neff(D=1)} except for the results in DWIA.}
\begin{tabular}{cccc}  \\ \hline\hline 
pion wave function & $N_{\rm eff}(1s) $ & $N_{\rm eff}(2s)$ &$N_{\rm
 eff}(1s)/N_{\rm eff}(2s)$  \\ \hline
$\phi^{R}_{l_\pi}(r)$ & $2.05 \times 10^{-2}$ & $4.09 \times 10^{-3}$ & 5.01 \\
$\phi_{l_\pi}(r)$ & $2.08 \times 10^{-2}$ & $4.17 \times 10^{-3}$ & 4.99 \\ \hline
\label{Neff(D)}
\end{tabular}
\end{center}
\end{table}

We show in Fig.~\ref{WF(ren)} the pion radial densities of 1$s$ and 2$s$ 
states in $^{121}$Sn calculated by $\phi^{R}$ and $\phi$ in
Eq.~(\ref{WF_ren}).
Because $Z^{*}_{\pi}$ is one outside the nucleus, 
the pion densities are modified only inside the nucleus 
$0 \leq r \lesssim 8$ fm as can be seen in Fig~\ref{WF(ren)}.
In Fig.~\ref {Cross-ren}, we show the contributions of
($3s_{1/2})^{-1}_{n}$ neutron-hole state to the 
$^{122}$Sn($d, ^{3}$He) spectra for the formation of pionic atoms 
and we find the effects due to $Z^{*}_\pi$ factor are tiny.
To estimate the effects of $Z^{*}_{\pi}$ to the observables, we show the
calculated effective numbers for the dominant subcomponents $[(1s)_{\pi}
\otimes (3s_{1/2})^{-1}_{n}]$ and $[(2s)_{\pi} \otimes
(3s_{1/2})^{-1}_{n}]$ in Tables~\ref{Neff(D=1)} and~\ref{Neff(D)}.
We find that the ratios $N_{\rm eff}(1s)/N_{\rm eff}(2s)$ of $N_{\rm eff}$ 
of pion 1$s$ and 2$s$ states
formation, which are expected to be good quantities to deduce the pion
properties independent on the uncertainties of the neutron wave function
as discussed in Appendix~\ref{AppendixC}, 
changes only around 1\% for PWIA and 0.4 \% for
DWIA for results with $\phi^R$ and $\phi$.
These numbers seem to be too small to observe experimentally at present.
Actually, the variation of the ratio of effective number 
$N_{\rm eff}(1s)/N_{\rm eff}(2s)$ due to the
uncertainties of neutron wave function is larger and is around 10\% as
discussed in Appendix~\ref{AppendixC}.
Hence, it seems difficult to deduce new information on
$Z^{*}_{\pi}$ from the observed spectra by simply using the ratio of 1$s$
and 2$s$ states formation strength.
We mention here that
the effects of $Z^{*}_{\pi}$ in the ratios of $N_{\rm eff}$ are
suppressed because the pionic 1$s$ and 2$s$ states probe the almost same
nuclear density as described in Section~\ref{Sec.3.2} and the
$Z^{*}_{\pi}$ effects for both states are cancelled out in the ratio.
The idea to extract new information on $Z^{*}_{\pi}$ from observables
are considered to be still relevant.

Finally, 
we also investigate the sensitivity of the neutron pick-up reactions to
the optical potential parameters which satisfy the Seki-Masutani
correlation Eq.~(\ref{SM1}).
Since the effective numbers calculated by Eq.~(\ref{Neff}) have
different $\phi_{l_{\pi}}$ dependence from the overlapping density
Eq.~(\ref{overlap}), we may have chance to distinguish the potential
parameters with Seki-Masutani correlation by the formation cross sections.
We consider three sets of $b_{0}$ and Re$B_{0}$ parameter which are,  
\{$b_0$, Re$B_0$\} = \{$-0.0185, -0.05$\}, 
 \{$-0.0300, 0.00$\}, and \{$-0.0415 , 0.05$\}, 
where parameters are in pion mass units,
and we find that there appear some discrepancies in the radial part of
the integrand of Eq.~(\ref{Neff}) around
$0 \leq r \lesssim 4.5$ fm.
Since the nuclear half radius is taken to be 5.4761 fm in this case, the
discrepancy only exists deep inside the nucleus which will be
significantly suppressed by the distortion factor.
Thus, we expect that the effects to the formation rate are extremely small.
We have confirmed the expectation by numerical calculation which show 
that the variation of the ratios of the effective numbers of 1$s$ and
2$s$ states for these potential parameter sets is less than 1 \%.

\section{Conclusion}\label{discus}
In this article, we have shown the newly calculated ($d, ^{3}$He)
spectra on the $^{122}$Sn,  $^{122}$Te, $^{126}$Te targets for the
formation of the pionic atoms~\cite{RIBF1,RIBF2}.
Based on these results,
we have investigated the possibilities to deduce the new information
on the $\langle {\bar q}q \rangle$ value at various nuclear densities and on
the wave function renormalization factor $Z^{*}_{\pi}$ paying attention to the
recent theoretical and experimental developments of the studies of the
pionic atoms.

We have found that the formation spectra on the $^{122}$Sn target is suited for
the observation of the pionic states because of the simple neutron level
structure and large occupation probability of $3s_{1/2}$ neutron state.
However, the ($d, ^{3}$He) reaction on Te isotopes could include the
extra difficulties due to the complex neutron level structure and should
be considered carefully for the experiments.

We have also found that the nuclear density probed by the atomic pion is
distributed only inside the narrow region around $\rho=0.6\rho_{0}$ even
for the deeply bound pionic states and the non-yrast states.
This feature had an advantage in early stage of the exploration, 
however, it requires now the experimental data with excellent precision
to deduce the information on $\langle {\bar q} q \rangle$ at various
nuclear density as shown in the contour plot studies in Section~\ref{Sec.3.2}.

As for the uncertainties of the calculated cross sections and the
determination of $Z^{*}_{\pi}$, we have found that the
ratio of the
subcomponents coupled with the same neutron hole state is a good index
relatively free from the systematic errors due to the neutron wave
function.
This observation is possible in experiments in RIBF/RIKEN.
However, the renormalization factor $Z^{*}_{\pi}$ changes the pion
wave function inside the nucleus slightly and 
the effects to the cross section are masked by the distortion factor.
Since the nuclear densities probed by atomic pions are around $0.6
\rho_0$, the effects of $Z^{*}_{\pi}$ for various pionic atoms are
almost same and are cancelled out in the ratio of the formation cross
sections.
Hence, the observation of $Z^{*}_{\pi}$ is rather difficult at preset.
However, the idea to deduce $Z^{*}_{\pi}$ information from observables
is important and the further studies are required.

We think that we need to consider the pionic atoms in exotic nuclei to
obtain the pion properties at various nuclear densities and the
information on pion wave function renormalization.
For example, the existence of the pionic nuclear states due to the
strong Coulomb attraction~\cite{Friedman} and due to the thick neutron
skin effects~\cite{PLB249} were predicted theoretically.
Pionic atoms in unstable nuclei were studied in various cases in
Refs.~\citen{PLB194,Fujita,RefJ8}.
In these systems with exotic nuclei, we may have the different
sensitivities of pions to nuclear densities and the different effects of
the renormalization factor to the formation spectra.
We will leave the study of these systems as future works.



\section*{Acknowledgements}
We would like to thank H. Toki, T. Yamazaki, R. S. Hayano and K. Suzuki
for many collaborations and fruitful discussions on the pionic atoms.
N. I. appreciates the support by the Grant-in-Aid for JSPS Fellows.
This work was partly supported by the Grants-in-Aid for Scientific Research 
(No.~22740161, No.~20540273, No.~22105510, and No.~22105517). 
This work was done in part under the 
Yukawa International Program for Quark-hadron Sciences (YIPQS).

\appendix
\section{Theoretical Formula for Structure and Formation of Pionic Atoms}
\label{struct-cross}
The theoretical formula to calculate the structure and formation of the
pionic atoms~\cite{RefJ4,RefJ5,RefJ6,RefJ7,RefJ11,RefJ9,RefJ10} are summarized in this Appendix.
The energy spectra and wave functions of the pionic atoms can be
obtained theoretically by solving the Klein-Gordon equation,
\begin{equation}
\left[-\nabla^{2}+\mu^{2}+2\mu V_{\rm{opt}}(r)\right]\phi({\boldsymbol r})
=\left[ E- V_{\rm{coul}}(r)\right]^{2} \phi({\boldsymbol r}),
\label{KGeq}
\end{equation}
where $\mu$ is the pion-nucleus reduced mass, $E$ the eigen energy
written as
$E=\mu-B_{\pi}-\displaystyle \frac{i}{2}\Gamma$ with the binding energy
$B_\pi$ and the width $\Gamma$ of the atomic states. 
The $V_{\rm coul}$ is the Coulomb potential with a finite nuclear charge
density distribution $\rho_{ch}(r)$:
\begin{equation}
V_{\rm{coul}}(r)
=-e^{2}\int\frac{\rho_{ch}(r^{\prime})}{|\boldsymbol {r}-\boldsymbol{
r^{\prime}}|}d \boldsymbol{r^{\prime}}.
\label{coulon-pot}
\end{equation}
The charge density distribution is written by the Woods-Saxon form as,
\begin{equation}
\rho_{ch}(r) = \frac{ \rho_{ch 0}}{1+ \exp[(r-R_{ch})/a_{ch}]}.
\end{equation}
The parameters of the charge distributions are taken from
Ref.~\citen{Fricke} and summarized in Table~\ref{R,a} for the nuclei
considered in this article.

The $V_{\rm opt}$ in Eq. (\ref{KGeq}) is the pion-nucleus optical potential,
which we assume to be of the Ericson-Ericson type~\cite{Ericson},
\begin{eqnarray}
2\mu V_{\rm opt}(r)
=&-&4\pi[b(r)+\varepsilon_2B_0\rho^2(r)]+4\pi\nabla\cdot[c(r)+
\varepsilon_2^{-1}C_0\rho^2(r)]L(r)\nabla,
\label{Vopt}
\end{eqnarray}
with
\begin{equation}
b(r)= \varepsilon_{1}[b_{0}\rho(r)+b_{1}[\rho_n(r)-\rho_p(r)]],
\label{Vopt_Swave1}
\end{equation}
\begin{equation}
c(r)=\varepsilon_{1}^{-1}[c_{0}\rho(r)+c_{1} [\rho_n(r)-\rho_p(r)]],
\label{Vopt_Pwave2}
\end{equation}
\begin{equation}
L(r)=\left\{ 1+ \frac{4}{3} \pi\lambda[c(r)+
      \varepsilon_{2}^{-1}C_{0}\rho^{2} (r)] \right\}^{-1}, 
\label{Vopt_Pwave3}
\end{equation}
where $\varepsilon_{1}$ and $\varepsilon_{2}$ are defined as
$\varepsilon_{1}=1+\displaystyle \frac{\mu}{M}$ and 
$\varepsilon_{2}=1+\displaystyle \frac{\mu}{2M}$ with 
the nucleon mass $M$.  
As a standard parameter set, we use the potential parameters
listed in Table~\ref{table:1}, which are taken from Ref.~\citen{SM}.
We use the Woods-Saxon form for the distributions of proton and neutron centers
and assume the same shape for the both distributions,
\begin{equation}
\rho(r)=\rho_{p}(r)+ \rho _{n}(r)
=\frac{ \rho_{0}}{1+ \exp[(r-R)/a]},
\end{equation}
where $R$ and $a$ are the radius and diffuseness parameters, 
which are determined from the parameters $R_{ch}$ and $a_{ch}$ of the
charge distributions $\rho_{ch}$ by the prescription described in
Ref.~\citen{oset}. 
For the calculations of the bound states, we use the distribution
parameters $R$ and $a$ same as the target nuclei of the ($d, ^3$He)
formation reaction.
The $\rho_0$ is obtained by the correct normalization of the mass number
of the daughter nucleus.

\begin{table}[tb]
\caption{Radius parameters $R_{ch}$ of the charge distributions
 taken from Ref~\citen{Fricke}. 
The diffuseness parameters are fixed to be $a_{ch}=t$/4ln3 for all
 nuclei with $t=2.30$ fm.}
\label{R,a}
\begin{center}
\begin{tabular}{c|cccccc} \hline \hline
nuclide & $^{116}$Sn & $^{120}$Sn & $^{122}$Sn & $^{124}$Sn &$^{122}$Te
 &$^{126}$Te \\ \hline
$R_{ch}$ [fm] & 5.4173 & 5.4588  & 5.4761 & 5.4907 & 5.5368 & 5.5617 \\ \hline
\end{tabular}
\end{center}
\end{table}
\begin{table}[htb]
\caption{Pion-nucleus optical potential parameters~\cite{SM} used in the
 present calculations.}
\label{table:1}
\begin{center}
\begin{tabular}{cc} \hline \hline
$b_{0} = -0.0283 m_{\pi}^{-1}$ &		
$b_{1} = -0.12   m_{\pi}^{-1}$ \\	
$c_{0} =  0.223  m_{\pi}^{-3}$ &	
$c_{1} =  0.25   m_{\pi}^{-3}$ \\	
$B_{0} =  0.042i m_{\pi}^{-4}$ &		
$C_{0} =  0.10i  m_{\pi}^{-6}$ \\  
$\lambda =  1.0$ \\ \hline			 
\end{tabular}
\end{center}
\end{table}

We can calculate the pionic atom formation cross sections 
in the effective number approach~\cite{RefJ6,RefJ7} using
the pionic atom wave function $\phi_{l_\pi}$, the binding energy
$B_{\pi}$, and the width $\Gamma$ obtained by solving the Klein-Gordon
equation Eq.~(\ref{KGeq}). 
The ($d$,$^3$He) reaction cross section in the laboratory frame is
expressed as, 
\begin{equation}
\left(\displaystyle{\frac{d\sigma}{d\Omega dE}}\right)_{dA\rightarrow^3
 {\rm{He}}(A-1)\pi} =\left(\displaystyle{\frac{d\sigma}{d\Omega}}\right)
 _{dn\rightarrow^3{\rm{He}}\pi} \times 
\sum_{[l_{\pi}\otimes j_n^{-1}]}\frac{\Gamma}{2\pi}\frac{1}{\Delta
E^{2}+{\Gamma}^{2}/4}
 N_{\rm{eff}},
\label{Cross}
\end{equation}
with 
\begin{eqnarray}
N_{\rm{eff}}=\sum_{JMm_{s}} &\Big|& \int d \boldsymbol{r}d\sigma
 \chi_{f}^{\ast}(\boldsymbol{r}) \xi^{\ast}_{1/2,m_{s}}(\sigma) 
 [\phi^{\ast}_{l_{\pi}}\otimes\psi_{j_{n}}(\boldsymbol{r},\sigma)]_{JM}\chi_{i}(\boldsymbol{r}) \Big|^{2}.
\label{Neff}
\end{eqnarray}
Here,
$\displaystyle \left(\frac{d\sigma}{d\Omega}\right) _{dn \rightarrow
^3{\rm{He}}\pi}$ 
indicates the elementary differential cross section at forward angles
for the $d+n \rightarrow {^3{\rm He}}+\pi^{-}$ reaction in the laboratory
system, which is extracted from the experimental data of the $p+d
\rightarrow \pi^{+}+ t $ reaction assuming charge symmetry\cite{RefJ6,RefJ7}.
$\Delta E$ is defined as $\Delta E=Q +m_{\pi}-B_{\pi}+S_{n}-6.787$
MeV for the ($d, ^3$He) reaction with the pion mass $m_{\pi}$, the pion
binding energy $B_{\pi}$, the
neutron separation energy $S_{n}$, and the reaction $Q$-value.
$\Gamma$ denotes the width of the bound pionic state.

For the neutron wave function $\psi_{j_n}$,
we adopt the harmonic-oscillator (HO) wave function in this article for
simplicity.
We also use the calculated neutron wave function with Woods-Saxon type 
potential to check the theoretical uncertainties of the cross sections
in Appendix~\ref{AppendixC}. 
We have used the oscillator parameter given
by $\hbar \omega=40A^{-\frac{1}{3}} $ MeV for the harmonic-oscillator
wave functions, with $A$ the nuclear mass number. 
The spin wave function is denoted as $\xi_{1/2,m_{s}}(\sigma)$, and we take
the spin average with respect to
$m_{s}$ so as to take into account the possible spin direction of the
neutrons in the target nucleus. 
$\chi_i$ and $\chi_f$ express the initial
and final distorted waves of the projectile and the ejectile,
respectively. We use the Eikonal approximation and replace $\chi_i$ and
$\chi_f$ according to
\begin{equation}
\chi_{f}^{*}(\boldsymbol{r})\chi_{i}(\boldsymbol{r})=\exp(i\boldsymbol{q}\cdot
 \boldsymbol{r})D(z, \boldsymbol{b}),
\label{D(z)}
\end{equation}
where the distortion factor $D(z, \boldsymbol{b})$  is defined as
\begin{eqnarray}
D(z, \boldsymbol{b})=\exp
 \Big[&-&\frac{1}{2}\sigma_{dN}\int_{-\infty}^{z}dz^{\prime}\rho_{A}(z^{\prime},\boldsymbol{b}) -
 \frac{1}{2}\sigma_{hN}\int_{z}^{\infty}dz^{\prime}\rho_{A-1}(z^{\prime},\boldsymbol{b})\Big]. 
\label{D(z)2}
\end{eqnarray}
Here, the deuteron-nucleon and $^3$He-nucleon total cross sections are
denoted as $\sigma_{dN}$ and $\sigma_{hN}$. 
The function $\rho_A(z,\boldsymbol{b})$ and
$\rho_{A-1}(z,\boldsymbol{b})$ are the density distributions of the
target and daughter nuclei at beam-direction coordinate $z$ with impact
parameter $\boldsymbol{b}$. 
The effective number approach is sometimes called the distorted wave
impulse approximation (DWIA), and also called the plane wave impulse
approximation (PWIA) in case we neglect the distortion effects in
Eqs.~(\ref{D(z)}) and (\ref{D(z)2}) as putting $D(z,\boldsymbol{b})=1$. 

In order to predict the spectrum of the ($d$,$^3$He) reactions, we need
to take into account the realistic ground-state configurations of the
target nuclei, the excitation energies, and the relative excitation
strengths leading to the excited states of the daughter nuclei. 
To obtain a realistic total
strength for the neutron pick-up from each orbital, we need to normalize
the calculated effective numbers using the neutron occupation
probabilities in the ground state of the target nucleus. The occupation
probabilities are obtained from the analyses of the $^{A}Z(d, t)^{A-1}Z$
reaction data and are not equal to one in general.

As for the excited levels of the daughter nuclei, we use the experimental
excitation energies and strengths obtained from the $^{A}Z(d,t)^{A-1}Z$
reaction. Since the single-neutron pick-up reaction from a certain
orbital in the target can couple to several excited states of the
daughter nuclei, we need to distribute the effective numbers among
these excited levels of the daughter nuclei in proportion to the
experimental strengths. Thus, the effective number for the pionic state
($\ell_\pi$) formation with the $N$-th daughter nucleus excited state coupled to
a single neutron pick-up from a neutron orbit $j_n$ is written 
\begin{equation}
N_{\rm{eff}}(\ell_{\pi}\otimes 
(\it{j_{n}}^{-1})_{N})=N_{\rm{eff}}(\ell_{\pi}\otimes\it{j_{n}}^{-1})
\times F_{O}(j_{n})
\times F_{R}((j_{n}^{-1})_{N}),
\label{Neff2}
\end{equation}
where $N_{\rm{eff}}(\ell_{\pi}\otimes \it{j_{n}}^{-1})$ is the effective
number defined in Eq. (\ref{Neff}), $F_{O}$ the normalization factor
due to the occupation probabilities of the neutron states $j_n$ in the
target nucleus, and $F_R$ is the relative strength of the $N$-th excited
states in the daughter nucleus coupled to the single neutron pick-up from
the state $j_n$.
The $F_{O}$ and $F_{R}$ of some medium
heavy nuclei are compiled in Tables \ref{Sn-Fo}, \ref{Sn-ExFr},
\ref{Te-Fo}, and \ref{Te-ExFr} given in Appendixes~\ref{AppendixC} and
\ref{AppendixD}, and also in Refs.~\citen{RefJ9,RefJ10}.

\section{Nuclear Structure Dependence of Formation Cross Section of 
Pionic Atoms}\label{AppendixC}
We discuss in this Appendix the uncertainties of the effective number
approach used to calculate the cross sections. 
As we can see from the data in Ref.~\citen{RefJ1} and the theoretical
prediction in Ref.~\citen{RefJ10}, the effective number approach works
well to predict the shape of the ($d, ^{3}$He) spectra, however, it
fails to predict the absolute magnitude of the cross sections correctly.
Thus, it is important to know the limitation of the applicability of
this approach.
This is also important to apply this approach to deduce $Z^{*}_{\pi}$
from experimental data as discussed in Section~\ref{Sec.3.3}.

\begin{figure}[tb]
\begin{minipage}[t]{0.48\hsize}
\begin{center}
\centerline{\includegraphics [width=6.6cm, height=5.3cm]{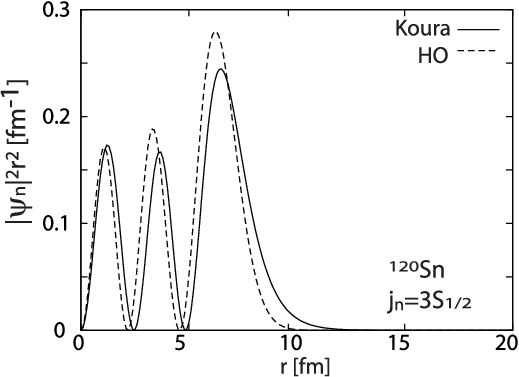}}
\caption{Neutron distributions of $j_{n}=3s_{1/2}$ state in $^{120}$Sn
by the harmonic oscillator wave function (dashed line) and the
calculated wave function by the potential (set OB) in Ref~\citen{koura}
(solid line).}
\label{HO_wave}
\end{center}
\end{minipage}
\begin{minipage}[t]{0.48\hsize}
\begin{center}
\centerline{\includegraphics [width=6.6cm, height=5.3cm]{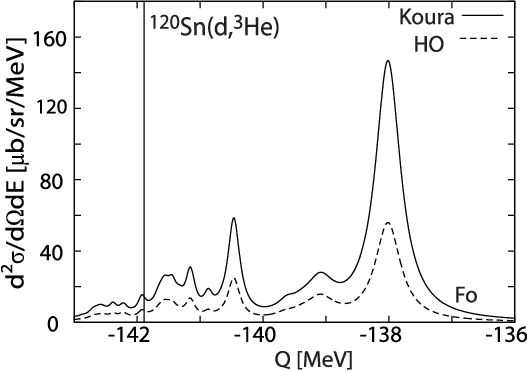}}
\caption{Expected spectra of the $^{120}$Sn($d, ^3$He) reaction for the
formation of deeply bound pionic atoms at $T_{d}=500$ MeV plotted as
functions of the reaction $Q$-value.
The harmonic oscillator (dashed line) and the theoretical (solid line)
neutron wave functions~\cite{koura} are used. 
Instrumental energy resolution is assumed to be 150 keV FWHM.
The vertical line indicates the threshold $Q=-141.9$ MeV.}
\label{HO_cross}
\end{center}
\end{minipage}
\end{figure}

We show first the dependence of the calculated cross sections 
on the neutron wave functions $\psi_{n}$ in targets.
As mentioned in Appendix~\ref{struct-cross}, we adopted the HO wave 
functions for simplicity in this article.
We calculate here the formation spectra using another set of $\psi_{n}$
obtained by a theoretical potential Set OB in Ref.~\citen{koura}.
The both neutron wave functions in $3s_{1/2}$ state in $^{120}$Sn are
shown in Fig.~\ref{HO_wave} as an example.
The $^{120}$Sn target nucleus was used in the latest experiment~\cite{RefJ1}
and considered in the theoretical calculations~\cite{RefJ9,RefJ10} before.
The $3s_{1/2}$ neutrons have dominant contributions to the cross section
coupled with pionic $s$-states.
We can see in Fig.~\ref{HO_wave} that the both wave functions show similar behavior, however
the wave function calculated by the potential in Ref.~\citen{koura} has
a little longer tail than that of the HO.
Because of the distortion effects, the long range tail part of the wave
function can be important for evaluating the formation rate.

Using these wave functions, we have calculated the $^{120}$Sn ($d, ^3$He)
spectra for the pionic atom formation and showed the results in Fig.~\ref{HO_cross}.
The $F_{O}$ and $F_{R}$ factors appeared in Eq.~(\ref{Neff2}) are taken
from Table~IV in Ref.~\citen{RefJ10}.
We found that the cross sections calculated with $\psi_{n}$ of
Ref.~\citen{koura} is about factor 3 larger than those with the harmonic
oscillator.
The theoretical calculations in Refs.~\citen{RefJ9,RefJ10}, which adopted the
same harmonic oscillator wave functions for $\psi_{n}$ with different
parameterization of proton and neutron density distributions in 
$V_{\rm opt}$, show the similar values for cross sections with the present case of
HO wave functions, while the experimental result in Ref.~\citen{RefJ1}
shows the smaller strength than theoretical calculations.
We think that the tail part of the wave function is significantly
important to evaluate the formation cross sections because the inner part
of the wave function is masked by the distortion effects.
And at the same time,  
it will be very difficult to obtain the tail of
$\psi_{n}$ precisely and to make the accurate predictions of the
absolute values of the formation cross sections.
On the other hand, the shape of the spectra is relatively robust and is
insensitive to $\psi_{n}$ as shown in Fig.~\ref{HO_cross}.
Thus, it is better to use the ratio of the formation rates of
1$s$ and 2$s$, for example, to deduce the information of $Z^{*}_{\pi}$ than
to use the absolute peak height of each state to reduce the
uncertainties due to $\psi_{n}$.
The variation of the ratio of the effective numbers for the
subcomponents $[(1s)_{\pi} \otimes (3s_{1/2})^{-1}_{n}]$ and 
$[(2s)_{\pi} \otimes (3s_{1/2})^{-1}_{n}]$ due to the different neutron wave
functions is around 10 \% for the case in Fig.~\ref{HO_cross}.

We then consider the uncertainties due to the neutron occupation
probability ($F_{O}$) in the target nucleus and the relative strength
of the excited levels ($F_{R}$) of the daughter nucleus. 
We summarize $F_{O}$ and $F_{R}$ together with the excited energies
($E_{x}$) in Table~\ref{Sn-Fo} and \ref{Sn-ExFr}.
$F_{R}$ and $E_{x}$ of $^{119, 123}$Sn can be found in Table IV of
Ref.~\citen{RefJ10}.
We show three sets of $F_{O}$ values in Table~\ref{Sn-Fo} to estimate the
uncertainties of the $F_{O}$ values.
Within these neutron states, it is known that the $s_{1/2}$ and $d_{5/2}$
states have dominant contributions to the ($d, ^{3}$He) spectra in
recoilless kinematics for the pionic atom formation.
As we can see from Table~\ref{Sn-Fo}, the calculated results by RMF
model with and without including the
possibility of nuclear deformation ($F^{d}_{O}$ and $F^{s}_{O}$) show
the qualitatively same results.
This fact indicates that these Sn isotopes are almost spherical nuclei.
In addition, $F_{O}$ obtained from experimental data also show similar
results.
On the other hand, 
the results of Te isotopes show significantly different and complicated
features as shown in Appendix~\ref{AppendixD}.
Thus,  
the uncertainties due to $F_{O}$ for Sn isotopes, which are roughly
20-30 \%, are much smaller than those for Te isotopes.
This ambiguity, however, does not affect the ratio between
subcomponents with the same neutron hole state.

As for the excited states of the daughter nuclei of the ($d, ^{3}$He)
reaction, the relative strength
$F_{R}$ is listed in Table~\ref{Sn-ExFr} for $^{121}$Sn and in Table IV
of Ref.~\citen{RefJ10} for $^{119, 123}$Sn.
It is difficult to evaluate the systematic errors of these numbers.
We can say, however, that the strengths of the excited states of Sn
isotopes are suited for the pionic atom formation because of 
their simplicity.
The only $2d_{5/2}$ excited level of $^{119, 121, 123}$Sn isotopes has a
little complex structure including a few levels, however, there is a
clear dominant level with the largest  value of $F_{R}$ even
for the $d_{5/2}$. 
On the other hand the excited levels in $^{121, 125}$Te include several
levels of the same quantum numbers with comparable strength as shown in
Appendix~\ref{AppendixD}, which makes
the reaction spectra rather complicated and prevents us from deducing the
information on pion.
Thus, we can say that the Sn isotopes are more suited as the targets of
the ($d, ^{3}$He) reactions for the formations of the pionic atoms.

\begin{table}[tb]
\caption{
Normalization factors, which correspond to the occupation probability
of each neutron state for the ground state of the target nucleus, are shown
for $^{120}$Sn, $^{122}$Sn, and $^{124}$Sn nuclei.
The normalization factors indicated by $F_{O}$ are evaluated by
experimental data.~\cite{RefJ10,Exp-Fo2} \ 
$F^{s}_{O}$ and $F^{d}_{O}$ are obtained by
the theoretical calculations of RMF model.~\cite{Gen} \ 
$F^{s}_{O}$ is calculated by assuming the spherical shape of nuclei,
and $F^{d}_{O}$ by including the deformation effects.
The numbers with asterisk for $1g_{7/2}$ and $1h_{11/2}$ states are
estimated so as to satisfy the normalization of total neutron number in
the valence shells by assuming the same occupation probabilities for
both states.}
\label{Sn-Fo} 
\begin{minipage}[t]{0.33\hsize} 
\begin{center} 
\begin{tabular}{p{10mm}p{6mm}p{6mm}p{6mm}}\hline\hline
$F_{O}$\\
Neutron orbit & $^{120}$Sn & $^{122}$Sn & $^{124}$Sn \\ \hline
$3s_{1/2}$ & 0.70 & 0.73 & 0.80 \\
$2d_{3/2}$ & 0.50 & 0.51 & 0.69 \\
$2d_{5/2}$ & 0.94 & 0.86 & 0.94 \\ 
$1g_{7/2}$ & 0.70 & $0.67^{*}$ & $0.70^{*}$ \\
$1h_{11/2}$ & 0.42 & $0.67^{*}$ & $0.70^{*}$ \\ \hline
\end{tabular}
\end{center}
\end{minipage}
\begin{minipage}[t]{0.33\hsize}
\begin{center}
\begin{tabular}{p{10mm}p{6mm}p{6mm}p{6mm}}\hline\hline
$F_{O}^{s}$ & & & \\
Neutron orbit & $^{120}$Sn & $^{122}$Sn & $^{124}$Sn \\ \hline
$3s_{1/2} $ & 0.54 & 0.65 & 0.78 \\ 
$2d_{3/2} $ & 0.64 & 0.77 & 0.85 \\ 
$2d_{5/2} $ & 0.93 & 0.95 & 0.96 \\ 
$1g_{7/2} $ & 0.95 & 0.98 & 0.97 \\ 
$1h_{11/2}$ & 0.28 & 0.33 & 0.47 \\ \hline
\end{tabular}
\end{center}
\end{minipage}
\begin{minipage}[t]{0.33\hsize}
\begin{center}
\begin{tabular}{p{10mm}p{6mm}p{6mm}p{6mm}}\hline\hline
$F_{O}^{d}$  \\ 
Neutron orbit & $^{120}$Sn & $^{122}$Sn & $^{124}$Sn \\ \hline
$3s_{1/2} $ & 0.53 & 0.67 & 0.78 \\
$2d_{3/2} $	& 0.68 & 0.76 & 0.85 \\
$2d_{5/2} $ & 0.94 & 0.95 & 0.96 \\
$1g_{7/2} $ & 0.97 & 0.97 & 0.97 \\
$1h_{11/2}$	& 0.24 & 0.35 & 0.45 \\ \hline
\end{tabular}
\end{center}
\end{minipage}
\end{table} 

\begin{table}[ht]
\caption{Excitation energy ($E_{x}$) and relative strength ($F_{R}$) of
 each excited level in $^{121}$Sn determined from the experimental data
of Ref.~\citen{Exp-Fo2}.}
\label{Sn-ExFr}
\begin{center}
\begin{minipage}{0.5\hsize}
\begin{tabular}{p{30mm}p{15mm}p{15mm}} \\ \hline\hline
{$^{121}\rm{Sn}$} & & \\ 
Neutron hole orbit & $E_x$ [MeV] & $F_{R}$ \\ \hline
$3s_{1/2}$  & 0.06 &  1.00 \\
$2d_{3/2}$  & 0.00 &  1.00 \\
$2d_{5/2}$  & 1.11 &  0.65 \\
            & 1.37 &  0.35 \\
$1g_{7/2}$  & 0.90 &  1.00 \\
$1h_{11/2}$ & 0.05 &  1.00 \\ \hline
\end{tabular}
\end{minipage}
\end{center}
\end{table}

\section{Structure of Pionic Atoms in $^{121}$Sn and the Effective Nuclear Density
 probed by Atomic Pion}
\label{AppendixB}
\begin{table}[tb]
\caption{Calculated binding energies B.E. and widths $\Gamma$ of
$\pi^{-}-$$^{121}$Sn atom in units of keV.
B.E.$_{\rm FC}$ indicates the binding energies calculated with a
finite-size Coulomb potential only. }
\label{table:2}
\begin{center}
\begin{tabular}{c|c|cc} \hline\hline
& \multicolumn{3}{|c}{$\pi^{-}-$$^{121}$Sn}  \\ \cline{2-4} 
\multicolumn{1}{c|} {state} & B.E.$_{\rm FC}$
[keV] & B.E. [keV] & $\Gamma$[keV] \\
\hline
1$s$ & 6227.7 &  3829.2   & 320.6  \\	   	   
2$s$ & 1913.9 &  1415.6   &  76.8  \\
3$s$ &  909.2 &   732.5   &  29.3  \\	 
4$s$ &  528.3 &   446.6   &  14.1  \\
5$s$ &  344.7 &   300.5   &   7.8  \\
6$s$ &  242.4 &   215.8   &   4.8  \\	 
2$p$ & 2321.2 &  2262.7   & 116.7  \\	   
3$p$ & 1034.9 &  1013.6   &  39.6  \\
4$p$ &  582.5 &   572.9   &  17.5  \\	   
5$p$ &  372.8 &   367.7   &   9.2  \\	   
6$p$ &  258.8 &	  255.8   &   5.4  \\	   	   
3$d$ & 1037.8 &  1040.5   &   2.5  \\	    
4$d$ &  584.0 &	  585.6   &   1.5  \\	   	   
5$d$ &  373.6 &   374.6   &   9.0$\times 10^{-1}$ \\	   
6$d$ &  259.3 &   259.9   &   5.6$\times 10^{-1}$ \\	   
4$f$ &  581.8 &   581.8   &   5.6$\times 10^{-3}$ \\	   
5$f$ &  372.5 &   372.5   &   4.8$\times 10^{-3}$ \\	   
6$f$ &  258.7 &	  258.7	  &   3.5$\times 10^{-3}$  \\ \hline	 
\end{tabular}
\end{center}
\end{table}

\begin{figure}[tb]
\centerline{\includegraphics [width=6.7cm, height=11.5cm]{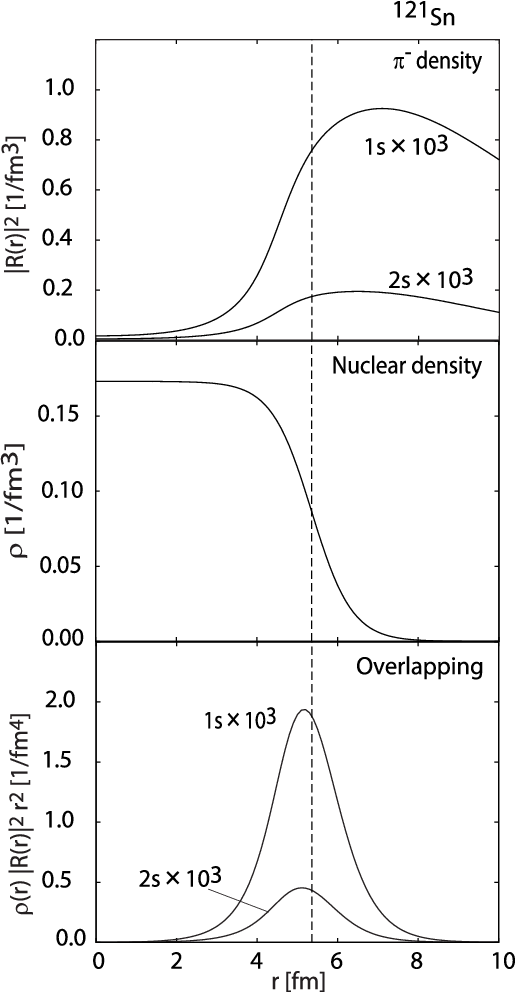}}
\caption{Overlapping densities (lower frame) of the $\pi^{-}$
densities (upper frame) with the nuclear density (middle frame) in
pionic bound 1$s$ and 2$s$ states in $^{121}$Sn. The vertical broken
line shows the half-density radius of the nuclear density of
$^{121}$Sn.}
\label{fig:2}
\end{figure}

We show the calculated binding energies and widths in
Table~\ref{table:2} for $\pi^{-}$ atoms in $^{121}$Sn nucleus, which
are expected to be observed in near future at
RIKEN /RIBF.\cite{RIBF1,RIBF2} \ 
We find, as the previous works~\cite{RefJ4,RefJ5,RefJ9,RefJ10} showed, that the
optical potential acts as
the soft repulsive core and moved pionic wave functions outwards.
This effect makes the widths of the bound states smaller to be 
quasi-stable states.
As we can see from Table~\ref{table:2}, the eigen energies of pionic
bound states are spread in wide region such as 250 $\lesssim$ B.E. $\lesssim$
3800 [keV] and $10^{-3} \lesssim \Gamma \lesssim 320$ [keV]
for the states with principal quantum number $n=1 \sim 6 $, and the shapes of
pion densities are changed significantly for different states.

We, then, consider 
the effective nuclear density probed by atomic pion, which was proposed
in Ref.~\citen{rho_e} and was one of
the good quantities to know the nuclear density sensitively observed by
pion in atomic states.
The definition of the effective nuclear density $\rho_e$ and the
overlapping density $S(r)$ are given in Section~\ref{Sec.3.2} and
Ref.~\citen{rho_e}.
We show as an example calculated pion densities $|R|^{2}$, nuclear
density $\rho$, and the overlapping densities $S$ for 1$s$ and 2$s$ states
in $^{121}$Sn$-\pi^{-}$ atoms in Fig.~\ref{fig:2}. 
We find the same tendency as found in Ref.~\citen{rho_e} for different
systems. 
Namely, the peak positions of the overlapping density $S(r)$
and, thus, the effective density
$\rho_{e}$ values are almost same for 1$s$ and 2$s$ atomic states even
their eigen energies and wave functions are much different.
We can interpret this tendency by considering the fact that
the repulsive optical potential and 
the attractive Coulomb potential make the potential pocket at nuclear
surface, where the overlapping densities are localized.

\section{Numerical results for Te isotopes}\label{AppendixD} 
\begin{table}[bt]
\caption{Calculated binding energies B.E. and widths $\Gamma$ of
$\pi^{-}-$$^{121}$Te and $\pi^{-}-$$^{125}$Te atoms in units of keV.}
\label{Te-BE}
\begin{center}
\begin{tabular}{c|cc|cc}\hline\hline
& \multicolumn{2}{|c} {$\pi^{-}-$$^{121}$Te}  
& \multicolumn{2}{|c} {$\pi^{-}-$$^{125}$Te}\\  
\multicolumn{1}{c|} {state} & B.E. [keV] & $\Gamma$[keV]& 
B.E. [keV] & $\Gamma$[keV] \\\hline
1$s$ & 4096.3  & 372.5  &  4037.1  & 342.4 \\	   	   
2$s$ & 1522.0  &  93.3  &  1507.3  &  84.1 \\
3$s$ &  789.1  &  36.2  &   783.4  &  32.5 \\	 
4$s$ &  481.6  &  17.6  &   478.9  &  15.7 \\
5$s$ &  324.2  &   9.8  &   322.7  &   8.8 \\
6$s$ &  233.0  &   6.0  &   232.1  &   5.4 \\	 
2$p$ & 2445.2  & 141.9  &  2432.7  & 133.7 \\	   
3$p$ & 1095.8  &  48.4   & 1091.4  &  45.4 \\
4$p$ &  619.4  &  21.5   &  617.5  &  20.1 \\	   
5$p$ &  397.6  &  11.3   &  396.6  &  10.5 \\	   
6$p$ &  276.7  &   6.6   &  276.1  &   6.2 \\	   	   
3$d$ & 1126.6  &   3.3   & 1126.7  &   3.4 \\	    
4$d$ &  634.1  &   2.0   &  634.1  &   2.1 \\	   	   
5$d$ &  405.5  &   1.2   &  405.6  &   1.2 \\	   
6$d$ &  281.4  &   7.5$\times 10^{-1}$  & 281.4  &  7.6$\times 10^{-1}$ \\	   
4$f$ &  629.5  &   8.0$\times 10^{-3}$   & 629.5 &  8.5$\times 10^{-3}$ \\	   
5$f$ &  403.0  &   6.9$\times 10^{-3}$  &  403.1 &  7.3$\times 10^{-3}$ \\	   
6$f$ &  279.9  &   5.1$\times 10^{-3}$  &  279.9 & 5.4$\times 10^{-3}$  \\ \hline	 
\end{tabular}
\end{center}
\end{table} 

We show in this Appendix that the calculated results for Te isotopes,
which are candidate nuclei of the future pionic atom experiments in
RIBF/RIKEN~\cite{RIBF1,RIBF2}.
In Table~\ref{Te-BE}, we show the calculated results of the binding
energies and widths of the pionic states in $^{121}$Te and $^{125}$Te.
We find that the level spacing is larger enough than the level widths
and the states are quasi-stable as other deeply bound pionic atoms.

\begin{table}[tb]
\caption{Normalization factors, which correspond to the occupation
 probability
 of each neutron state for the ground state of the nucleus, are shown
 for $^{122}$Te and $^{126}$Te nuclei.
The normalization factors indicated by $F_{O}$ are evaluated by
 experimental data.~\cite{Exp-Fo} \ 
The normalization factors $F^{s}_{O}$ and $F^{d}_{O}$ are obtained by
 the theoretical calculations of RMF model.~\cite{Gen} \ 
The factor $F^{s}_{O}$ is calculated by assuming the spherical shape of nuclei,
and $F^{d}_{O}$ by including the deformation effects.}
\label{Te-Fo}
\begin{minipage}[t]{0.33\hsize}
\begin{center}
\begin{tabular}{p{10mm}ll} \\ \hline\hline
$F_{O}$\\ 
Neutron orbit & $^{122}$Te &$^{126}$Te	\\ \hline
$3s_{1/2} $ & 0.34  & 0.50 \\
$2d_{3/2} $ & 0.31  & 0.47 \\
$2d_{5/2} $ & 0.65  & 1.00 \\
$1g_{7/2} $ & 0.43  & 0.59 \\
$1h_{11/2}$ & 0.23  & 0.42 \\ \hline
\end{tabular}
\end{center}
\end{minipage}
\begin{minipage}[t]{0.33\hsize}
\begin{center}
\begin{tabular}{p{10mm}rrrr} \\ \hline\hline
$F_{O}^{s}$& &  & \\
Neutron orbit &  $^{122}$Te &$^{126}$Te	\\ \hline
$3s_{1/2} $  & 0.41  & 0.69 \\
$2d_{3/2} $  & 0.58  & 0.80 \\
$2d_{5/2} $  & 0.90  & 0.95 \\
$1g_{7/2} $  & 0.96  & 0.97 \\
$1h_{11/2}$  & 0.33  & 0.51 \\ \hline
\end{tabular}
\end{center}
\end{minipage}
\begin{minipage}[t]{0.33\hsize}
\begin{center}
\begin{tabular}{p{10mm}rrr} \\ \hline\hline
$F_{O}^{d}$ &    &    &        \\ 
Neutron orbit &  $^{122}$Te & $^{126}$Te	\\ \hline
$3s_{1/2} $  & 0.02  & 0.05	\\
$2d_{3/2} $  & 0.36  & 0.85	\\
$2d_{5/2} $  & 0.95  & 0.98	\\
$1g_{7/2} $  & 0.98  & 0.99	\\
$1h_{11/2}$  & 0.41  & 0.56	\\ \hline
\end{tabular}
\end{center}
\end{minipage}
\end{table} 

\begin{table}[t]
\caption{Excitation energy ($E_{x}$) and relative strength ($F_{R}$) of
 each excited level in $^{121}$Te and  $^{125}$Te determined from the
 experimental data of Ref.~\citen{Exp-Fo}.}
\label{Te-ExFr}
\begin{minipage}[t]{0.46\hsize}
\begin{center}
\begin{tabular}{p{30mm}p{10mm}p{10mm}} \\ \hline\hline
{$^{121}\rm{Te}$} &    &             \\ 
Neutron hole orbit   & $E_x$[MeV] & $F_{R}$  \\ \hline
$3s_{1/2}$ 	& 0.00 &  1.00 \\
$2d_{3/2}$	& 0.21 &  1.00 \\
$2d_{5/2}$ 	& 0.48 &  0.37 \\
            & 0.59 &  0.28 \\
            & 0.92 &  0.09 \\
            & 1.17 &  0.15 \\
            & 1.32 &  0.05 \\
            & 1.36 &  0.06 \\
$1g_{7/2}$  & 0.45 &  0.71 \\
            & 1.15 &  0.29 \\
$1h_{11/2}$	& 0.29 &  1.00 \\ \hline
\end{tabular}
\end{center}
\end{minipage}
\begin{minipage}[t]{0.46\hsize}
\begin{center}
\begin{tabular}{p{30mm}p{10mm}p{10mm}} \\ \hline\hline
{$^{125}\rm{Te}$}  \\
Neutron hole orbit & $E_x$[MeV] & $F_{R}$ \\ \hline
$3s_{1/2}$ & 0.00 &  1.00 \\
$2d_{3/2}$ & 0.04 &  1.00 \\
$2d_{5/2}$ & 0.64 &  0.35 \\
           & 1.05 &  0.19 \\
           & 1.14 &  0.19 \\
           & 1.27 &  0.15 \\
           & 1.43 &  0.11 \\
$1g_{7/2}$ & 0.64 &  0.73 \\
           & 1.74 &  0.27 \\
$1h_{11/2}$& 0.00 &  1.00 \\ \hline
\end{tabular}
\end{center}
\end{minipage}
\end{table}

\begin{figure}[tb]
\begin{center}
\includegraphics [width=13.2cm,height=5.3cm]{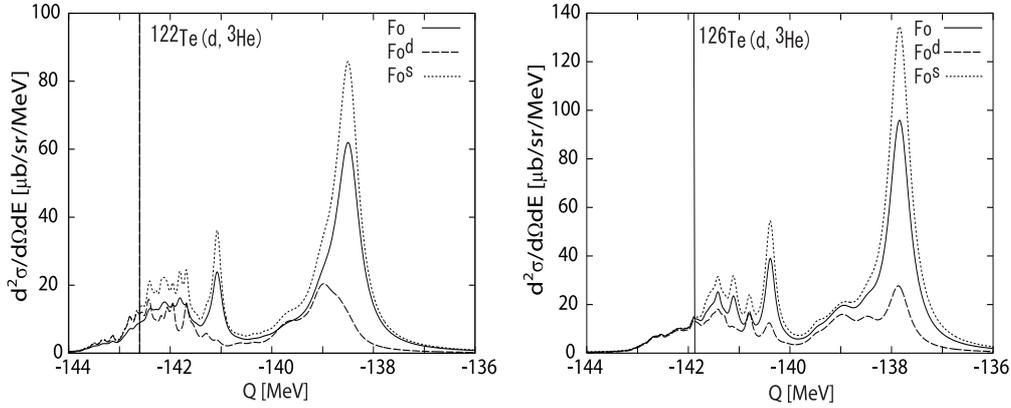}
\caption{Calculated $^{122}$Te($d, ^{3}$He) (left) and
 $^{126}$Te($d, ^{3}$He) (right) spectra for the formation of the pionic
 bound states at $T_{d}=500$MeV are shown as functions of the reaction
 $Q$-value.
Each line indicates the result with the different normalization factors
 ($F_{O}, F^{s}_{O}, F^{d}_{O}$) shown in Table~\ref{Te-Fo}.
The vertical line indicated the threshold $Q=-142.6$ MeV (left) and $-141.9$
 MeV (right).}
\label{cross_Te3}
\end{center}
\end{figure}

In Table~\ref{Te-Fo}, we show the occupation probabilities ($F_{O}$) of
the target nuclei $^{122}$Te and $^{126}$Te.
The factor $F_{O}$ evaluated with the experiment is listed together with
theoretical values, $F^{s}_{O}$ and $F^{d}_{O}$.
As you can see in the tables, the occupation of $3s_{1/2}$ is
significantly depend on the evaluation method.
Since $F^{d}_{O}$ is much smaller than $F^{s}_{O}$ for the
$3s_{1/2}$ level, we can expect
that both $^{122}$Te and $^{126}$Te are largely deformed. 
However, the experimental $F_{O}$ based on Ref.~\citen{Exp-Fo} are
closer to the spherical nuclear value $F^{s}_{O}$.
This feature seems somehow inconsistent, and these numbers could include
large errors. 
Actually larger deformations for Te isotopes than Sn isotopes were
reported in Ref.~\citen{ADNDT}.
In Table~\ref{Te-ExFr}, we show the relative strength $F_{R}$ for
excited levels of daughter nuclei $^{121}$Te and $^{125}$Te.
Here, we can see that the level structures of both nuclei are a little
complicated. 
Especially, the $2d_{5/2}$ state splits into several levels, which
include plural levels with similar strength.
These structures may cause extra difficulties to deduce the pion
properties from the formation spectra of the pionic atoms.

We show in Fig.~\ref{cross_Te3} the calculated ($d, ^{3}$He) spectra for
the $^{122}$Te and $^{126}$Te targets.
We find that the shape of the calculated spectra strongly depends on the
choice of $F_{O}$ as we expected.
The results with $F_{O}$ and $F^{s}_{O}$ show the qualitatively same
behavior, however, the results with $F^{d}_{O}$ show the much
different behavior.
This is due to the smallness of $F^{d}_{O}$ for $3s_{1/2}$ neutron
state, which can have dominant contribution to the spectra coupled with the
pionic $s$-states in the recoilless kinematics.
The lack of the $3s_{1/2}$ neutron contribution deformed the shape
of the spectra drastically.
In the spectra calculated with $F^{d}_{O}$,
the contributions of many subcomponents
compose the total spectra and, hence, they make it difficult to deduce
pion information clearly from the total spectra in this case.

Thus, we can conclude here that the expected spectra of the ($d,
^{3}$He) reaction on the Te isotope targets for the pionic atom
formation mentioned in Ref.~\citen{RIBF1} could include large
uncertainties due to nuclear structure which should be considered
carefully before experiments.

%

\end{document}